\documentclass[12pt,a4paper]{article}

\newcommand{\la}{\lambda}
\newcommand{\pa}{\partial}
\newcommand{\mt}{\mathcal{T}}
\newcommand{\me}{\mathcal{E}}

\begin{document}

\begin{flushright}
hep-th/0409217
\end{flushright}
\vspace{1.8cm}

\begin{center}
 \textbf{\Large Circular and Folded Multi-Spin Strings \\
in Spin Chain Sigma Models}
\end{center}
\vspace{1.6cm}
\begin{center}
 Shijong Ryang
\end{center}

\begin{center}
\textit{Department of Physics \\ Kyoto Prefectural University of Medicine
\\ Taishogun, Kyoto 603-8334 Japan}  \par
\texttt{ryang@koto.kpu-m.ac.jp}
\end{center}
\vspace{2.8cm}
\begin{abstract}
From the SU(2) spin chain sigma model at the one-loop and two-loop orders
we recover the classical circular string solution with two  $S^5$ spins 
$(J_1, J_2)$ in the $AdS_5 \times S^5$ string theory. 
In the SL(2) sector of the one-loop spin chain sigma model 
we explicitly construct a solution which
corresponds to the folded string solution with one $AdS_5$ spin $S$
and one $S^5$ spin $J$. In the one-loop general sigma 
model we demonstrate that there exists a
solution which reproduces the energy of the circular constant-radii string
solution with three spins $(S_1, S_2, J)$.

\end{abstract} 
\vspace{3cm}
\begin{flushleft}
September, 2004
\end{flushleft}

\newpage
\section{Introduction}

The AdS/CFT correspondence \cite{MGW} has more and more revealed the deep
relations between the conformal gauge theory and the string theory in 
curved spaces. The solvability of the string theory in the pp-wave 
background \cite{RM,MT} has opened a new celebrated step to give an
interesting proposal identifying nearly point-like string states with
gauge invariant near-BPS operators with large R-charge in the BMN limit
for the $\mathcal{N}=4$ SU(N) super Yang-Mills (SYM) theory \cite{BMN}.
Various semiclassical extended string configurations with several large
angular momenta in $AdS_5 \times S^5$ which usually go beyond the BMN
scaling have been constructed extensively to study the AdS/CFT 
correspondence for non-BPS states \cite{GKP,FT,JR,JM,SFT,FAT,SAA,
AFR,ART} and reviewed in \cite{AAT}.

In the planar $\mathcal{N}=4$ SYM theory there has been an important
observation that the dilatation operator for the SO(6) sector can be
interpreted as a Hamiltonian of an integrable spin chain in the
one-loop approximation \cite{MZ,NBS}. The planar integrability has
been studied at higher loops \cite{BKS,NB}. It plays 
important roles for resolving the complicated mixing problem and
diagonalizing the dilatation operator in order to obtain the anomalous
dimensions of the gauge invariant conformal composite operators by using
the powerful method of the Bethe ansatz. Within a closed SU(2) sector
consisting of operators of two out of the three complex scalar fields,
the one-loop Bethe ansatz of \cite{MZ} has been extended to three 
loops in \cite{SS} using the three-loop integrability of \cite{BKS}.
An all-loop Bethe ansatz has been proposed and refined in \cite{BDS}.

The Bethe equation for the one-loop dilatation operator in the SU(2)
sector has been solved for the states that are dual to the folded 
and circular strings having two large angular momenta in 
$S^5$ \cite{BMS,BFS} to show that its solutions match with the
one-loop semiclassical predictions in \cite{SFT,SAA}.
At the one-loop and two-loop orders there has been a general
proof of the equivalence between the solutions of the Bethe equations in
the thermodynamical limit and the classical solutions of the string theory
for large conserved charges in the SU(2) sector \cite{KMM}.
Attempts to relate the higher conserved charges on both sides to each 
other have been made in \cite{AS,JE,KMM}, indicating that the integrable
structures on both sides of the correspondence are closely related.
Further important works on the ``spinning limit" of AdS/CFT have been
provided in \cite{AKL}, and various investigations of the gauge/string
duality have been presented in \cite{BGK,BS,WW}.

There arises a natural question about the more direct relation of the 
integrable spin chain system to the string theory.  The 
SU(2) (Heisenberg $\mathrm{XXX}_{1/2}$) spin chain with the Hamiltonian
given by the one-loop dilatation operator, in the limit of large
number of chain sites, has been described by an effective two-dimensional
ferromagnetic sigma model for a coherent-state expectation value of
the spin operator, which on the other hand precisely agrees with
the sigma model obtained from the rotating string with two spins in
$S^5$ by taking some adequate angular momentum limit in the 
$AdS_5 \times S^5$ string action \cite{MK}.
It has been shown that this one-loop identification at the level of 
actions extends to two loops within the SU(2) sector \cite{KRT}.
The derivation of the low energy effective sigma model from the spin
chain system to leading order requires considering only long 
wavelength spin configurations, and beyond one loop involves
quantum corrections from short wavelength modes.
From the string theory the same effective sigma model has been derived by
choosing a special gauge with a non-diagonal world-sheet metric and
implementing an order by order redefinition of the relevant field to
get an action linear in the time derivative.
The derivation of the SU(2) continuum spin chain sigma model interpolating
between the SU(2) ferromagnetic spin chain and the $AdS_5 \times S^5$
string theory has been extended to the SU(3) sector 
(with three spins in $S^5$) \cite{HL,ST}, and the SL(2) sector
(with one spin in $AdS_5$ and one spin in $S^5$) \cite{ST}, where
a general effective sigma model action is further derived 
by rearranging the $AdS_5 \times S^5$ string action in the large
$S^5$ spin limit, which generalizes the one in \cite{MK,KRT}
from the two-spin $(J_1,J_2)$ configuration to the more general 
configuration with two $AdS_5$
spins and three $S^5$ spins. There has been a construction of a spin
chain sigma model interpolating between the string theory on 
$AdS_5 \times S^5/Z_M$ and $\mathcal{N}=0,1,2$ orbifold field theories
originating from $\mathcal{N}=4$ SYM \cite{KI}.

The one-loop spin chain sigma model for the SU(2) sector has been solved
to present a solution which reproduces the first-order correction in the
expansion of the classical energy for the folded string solution with
two spins in $S^5$ or the corresponding one-loop term in the anomalous
dimension found by using the Bethe ansatz \cite{MK}. 
This solution has been generalized \cite{DR} to be related to a general
ansatz for rotating strings with two spins which can be reduced to the
Neumann-Rosochatius integrable system \cite{ART}.
The two-loop part of the spin chain sigma model has been also shown 
to reproduce precisely the second-order energy correction for the folded
string solution \cite{KRT}. The equations of motion for 
the one-loop spin chain sigma model in 
the SU(3) sector have been explicitly
solved \cite{HL} to recover the first-order energy correction for the 
circular, rational string solution rotating along three orthogonal 
directions of $S^5$, which is the constant-radii solution of the
Neumann-Rosochatius integrable system \cite{ART}.
The other solution has been presented \cite{KM} and shown to reproduce
the circular, elliptic three-spin $(J_1,J_2,J_3)$ string 
solution \cite{AFR,CK}. There has been a construction of a general
effective one-loop sigma model action with 8-dimensional target space
which agrees with a limit of the $AdS_5 \times S^5$ phase-space string
action, from which the pulsating solution \cite{JM} and its 
generalization have been reproduced \cite{MKT}.

We will extend the effective sigma model analysis to other classes of
configurations to get a better understanding of the mapping between the 
classical string theory and the quantum gauge theory. In the continuum
SU(2) spin chain sigma model including the first two- ``one-loop"
and ``two-loop"- terms, we will search for a solution which corresponds to
the circular string solution with two spins $(J_1,J_2)$ in $S^5$.
We will construct a solution for the SL(2) sector of the one-loop
effective sigma model and show how it reproduces the two-spin 
$(S,J)$ folded  string solution in the $AdS_5 \times S^5$ string theory.
Manipulating the general effective sigma model action we will present
a solution which corresponds to the circular constant-radii solution
with two spins in $AdS_5$ and one spin in $S^5$.

\section{The circular two-spin $(J_1,J_2)$  string solution}

The low-energy effective action of the SU(2) ferromagnetic spin chain with
the Hamiltonian given by the sum of the one-loop \cite{MK} and 
two-loop \cite{KRT} dilatation operators of the $\mathcal{N}=4$ SYM theory
was given by the expression of the perturbative expansion in the effective
coupling constant $\tilde{\la}=\la/J^2 =g_{YM}^2N/J^2$
\begin{equation}
S = S_{WZ} - J\int d\mathrm{t}
 \int_{0}^{2\pi} \frac{d\sigma}{2\pi}\left\{
\frac{1}{8}(\pa_{\sigma}n^i)^2 - \frac{\tilde{\la}}{32} \left[ 
(\pa_{\sigma}^2n^i)^2 - \frac{3}{4} (\pa_{\sigma}n^i)^4 \right]
+ O(\pa_{\sigma}^6n^i) \right \},
\label{stn}\end{equation}
where $S_{WZ}$ is the Wess-Zumino term and $J$ is the length of the spin
chain. The subleading $(\pa_{\sigma}^2n^i)^2$ term is given by a naive
continuum limit of the expectation value of the two-loop dilatation 
operator in a coherent state, while the $(\pa_{\sigma}n^i)^4$ term is
derived by taking the continuum limit of the quantum correction to the
discrete spin chain theory. The coherent state in the SU(2) ferromagnetic
spin chain is specified by a unit vector 
\begin{equation}
n_i = (\sin \theta \cos\phi, \sin\theta \sin\phi, \cos\theta).
\label{ni}\end{equation}

Alternatively the same effective two-dimensional sigma model action was
derived by a suitable limit of the string action in the 
$R\times S^3 \subset AdS_5 \times S^5$ space-time with metric
\begin{eqnarray}
ds^2 &=& -dt^2 + |dX_1|^2 + |dX_2|^2 = -dt^2 + d\psi^2 + 
\cos^2\psi d\varphi_1^2 + \sin^2\psi d\varphi_2^2 , \nonumber \\
X_1 &\equiv& X_1 + iX_2 = \cos\psi e^{i\varphi_1}, \hspace{1cm} 
X_2 \equiv X_3 + iX_4 = \sin\psi e^{i\varphi_2},
\end{eqnarray}
where the range of the angles is $0<\psi\le 2\pi, 0<\varphi_1 \le \pi,
0<\varphi_2 \le \pi$. The string states 
rotate in the two orthogonal planes
$(X_1,X_2)$ and $(X_3,X_4)$ with two angular momenta $(J_1,J_2)$.
The three-sphere is parametrized in terms of $CP^1$ coordinates 
$U_r (r=1,2)$ and a U(1) angle $\alpha$ as $X_1 = U_1e^{i\alpha},
X_2 = U_2e^{i\alpha}$ with $U_1=\cos\psi e^{i\beta}, 
U_2=\sin\psi e^{-i\beta}, \alpha = (\varphi_1 + \varphi_2)/2,
\beta = (\varphi_1 - \varphi_2)/2$. The common phase $\alpha$ represents a
simultaneous rotation in the two plane so that it corresponds to the total
spin $J = J_1 + J_2 \equiv \sqrt{\la}\mt$. Introducing a unit vector
$n_i$ belonging to a two-sphere instead of $U_r$ as
$n_i \equiv U^{\dagger}\sigma_i U, \; U = (U_1,U_2)$ where $\sigma_i$
are Pauli matrices, we have
\begin{eqnarray}
n_i &=& (\sin2\psi \cos2\beta,\; \sin2\psi \sin2\beta,\; \cos2\psi), 
\label{nis} \\
(ds^2)_{S^3} &=& (D\alpha)^2 + \frac{1}{4}dn_i^2 =(d\alpha + C)^2 +
d\psi^2 + \sin^22\psi d\beta^2, \; C = \cos 2\psi d\beta 
\end{eqnarray}
and the string action is expressed in terms of $t, \alpha, n_i$.
The ``longitudinal" coordinates $t$ and $\alpha$ are so eliminated through
the constraints that the resulting sigma model action is described by a
``transverse"  coordinate $n_i$ in the same form as 
(\ref{stn}). The leading term quadratic in derivatives in (\ref{stn}) is
derived in the usual conformal gauge, while in order to obtain the 
subleading term quartic in derivatives we use a non-diagonal uniform
gauge and make a systematic field redefinition order by order 
in $1/\mt$ for elimination of time derivatives \cite{KRT}.
We rescale the time coordinate t as $\mathrm{t} \rightarrow \tilde{\la}
\mathrm{t}$ in the effective action (\ref{stn}) and use the coordinate 
$x = J\sigma/2\pi$ to have
\begin{equation}
S = S_{WZ} - \int d\mathrm{t} \int_0^J dx \left\{
\frac{\la}{32\pi^2}(\pa_xn^i)^2 - \frac{\la^2}{512\pi^4} \left[ 
(\pa_x^2n^i)^2 - \frac{3}{4} (\pa_xn^i)^4 \right] \right \}.
\label{stx}\end{equation}

Now from the sigma model action (\ref{stx}) we are ready to construct
a solution which corresponds to a circular string configuration with
two spins $J_1$ and $J_2$. At lowest order the equations of motion for
the first and second terms in (\ref{stx}) give a solution
\begin{equation}
\phi = \omega \mathrm{t}, \hspace{1cm} \theta = \theta(x)
\label{pth}\end{equation}
with
\begin{equation}
 \pa_x \theta = \sqrt{a_0 + b_0\cos\theta},
\label{xth}\end{equation}
where $\theta$ and $\phi$ are polar coordinates for the unit vector $n_i$
in (\ref{ni}). The integration of (\ref{xth}) is expressed as
\begin{equation}
x = \int_0^{\theta} \frac{d\theta}{\sqrt{a_0 + b_0\cos\theta}},
\label{xin}\end{equation}
where the integration constant is chosen as $\theta =0$ at $x = 0$.
Here we are interested in the  $a_0 > |b_0|$ case which corresponds to the
circular two-spin configuration, and perform the integration 
\begin{equation}
\sin\frac{\theta(x)}{2} = \mathrm{sn}\left( \frac{\sqrt{a_0 + b_0}}{2}x,
k \right), \hspace{1cm} k = \sqrt{\frac{2b_0}{a_0 + b_0}}.
\label{snk}\end{equation} 
By comparing (\ref{ni}) with (\ref{nis}) we note that $\theta = 2\psi$
and then the range of $\theta$ is $0< \theta \le 4\pi$.
Owing to the fundamental period $4K(y_0)$ of sn where $K(y_0)$ is the 
complete elliptic integral of the first kind, we extract a relation
\begin{equation}
\frac{\sqrt{a_0 + b_0}}{2}J = 4K(y_0), \hspace{1cm} y_0 = k^2 
=\frac{2b_0}{a_0 + b_0}
\end{equation}
from (\ref{snk}) with $\theta(J) = 4\pi$ and $\mathrm{sn}(4K,k)=0$,
where the string wraps completely around a great circle.
Alternatively the total angular momentum or the length of the
spin chain is also obtained from (\ref{xin}) as
\begin{equation}
J = J_1 + J_2 = 4\int_0^{\pi} \frac{d\theta}{\sqrt{a_0 + b_0\cos\theta}}
= \frac{8K(y_0)}{\sqrt{a_0 + b_0}}.
\label{jyk}\end{equation}
The total component of the spin in the $z$ direction is 
expressed as
\begin{eqnarray}
S_3 &=& \frac{J_2 - J_1}{2} = - \frac{1}{2}\int_0^J dx \cos\theta
= -2\int_0^{\pi} d\theta \frac{\cos\theta}{\sqrt{a_0 + b_0\cos\theta}}
\nonumber \\
&=& -\frac{4[(a_0 + b_0)E(y_0) - a_0K(y_0)]}{b_0\sqrt{a_0 + b_0}},
\label{sy}\end{eqnarray}
where $E(y_0)$ is the complete elliptic integral of the second kind.
From(\ref{stx}) the energy of this solution is provided by
\begin{eqnarray}
E_1 &=& \frac{\la}{32\pi^2}\int_0^J dx \{(\pa_x\theta)^2 +
\sin^2\theta(\pa_x\phi)^2 \} =  \frac{\la}{32\pi^2}4\int_0^{\pi}d\theta
\frac{a_0 + b_0\cos\theta}{\sqrt{a_0 + b_0\cos\theta}}
\nonumber \\ 
&=& \frac{\la}{4\pi^2}\sqrt{a_0 + b_0}E(y_0).
\label{ey}\end{eqnarray}
Combining (\ref{ey}) with (\ref{jyk}) we have
\begin{equation}
E_1 = \frac{2\la}{\pi^2J}K(y_0)E(y_0)
\label{eke}\end{equation}
and the expression (\ref{sy}) with (\ref{jyk}) gives
\begin{equation}
\frac{J_2}{J} = \frac{1}{y_0} - \frac{E(y_0)}{y_0K(y_0)}.
\label{jjy}\end{equation}

In ref. \cite{BFS} by the semiclassical string approach the 
circular two-spin  string solution in $AdS_5 \times S^5$ was found and
characterized by a system of two transcendental equations
\begin{eqnarray}
\left(\frac{\me}{K(y)}\right)^2 &-& \left(\frac{y\mt_1}
{E(y)-(1-y)K(y)} \right)^2 = \frac{4}{\pi^2},
\label{tre} \\
\left(\frac{y\mt_2}{K(y)-E(y)}\right)^2 &-& \left(\frac{y\mt_1}
{E(y)-(1-y)K(y)} \right)^2 = \frac{4}{\pi^2}y,
\label{trt}\end{eqnarray}
where $\me = E/\sqrt{\la},\; \mt_1 = J_1/\sqrt{\la}, \; \mt_2 = 
J_2/\sqrt{\la}$. These equations were solved by using the expansions of
$y = y_0 + y_1/\mt^2 + y_2/\mt^4 + \cdots$ and $\me = \mt +
\epsilon_1/\mt + \epsilon_2/\mt^3 + \cdots, \mathrm{i.e.}\;  E = J + 
\epsilon_1\la/J + \epsilon_2\la^2/J^3 + \cdots$, and the following 
solution was presented:
\begin{equation}
\epsilon_1 = \frac{2}{\pi^2} K(y_0)E(y_0), \hspace{1cm}
\frac{J_2}{J} = \frac{1}{y_0} - \frac{E(y_0)}{y_0K(y_0)}.
\label{jyj}\end{equation}
The energy expression (\ref{eke}) obtained from the one-loop effective
Hamiltonian of the spin chain sigma model agrees with the leading
energy orrection $\epsilon_1\la/J$ and the result (\ref{jjy}) is the
same as the momentum fraction $J_2/J$ in (\ref{jyj}).  

Following the prescription in ref. \cite{KRT}, we estimate the
subleading energy correction. We perturb the leading-order solution
(\ref{pth}) with (\ref{xth}) keeping $J$ and $S_3$ fixed so that
the energy correction is provided by the evaluation of the subleading
term quartic in derivatives in (\ref{stx}) on the leading-order solution.
The subleading energy correction given by
\begin{equation}
E_2 = -\frac{\la^2}{512\pi^4} \int_0^J dx \left[ (\pa_x^2\theta)^2 +
\frac{1}{4}(\pa_x\theta)^4 \right]
\end{equation}
is evaluated on the unperturbed leading-order solution as
\begin{eqnarray}
E_2 &=& -\frac{\la^2}{512\pi^4} 4\int_0^{\pi} \frac{d\theta}{\pa_x\theta}
\left[ (\pa_x^2\theta)^2 + \frac{1}{4}(\pa_x\theta)^4 \right]
\nonumber \\
&=& -\frac{\la^2}{512\pi^4} \int_0^{\pi}d\theta \frac{a_0^2 + b_0^2 + 
2a_0b_0\cos\theta}{\sqrt{a_0 + b_0\cos\theta}} \nonumber \\
&=& \frac{2}{\pi^4}\frac{\la^2}{J^3}K^3(y_0)\{(1-y_0)K(y_0)
- (2-y_0)E(y_0) \}.
\label{etk}\end{eqnarray}
The eq. (\ref{trt}) is written by
\begin{equation}
\left(\frac{\mt_2/\mt}{K(y)-E(y)}\right)^2 - \left(\frac{\mt_1/\mt}
{E(y)-(1-y)K(y)} \right)^2 = \frac{4}{\pi^2y}\frac{1}{\mt^2},
\end{equation}
which can be expanded in $1/\mt^2$ by using the formulae
\begin{eqnarray}
K'(y_0) &=& \frac{E_0 - (1-y_0)K_0}{2y_0(1-y_0)}, \;
K''(y_0) = \frac{K_0}{4y_0(1-y_0)}- \frac{1-2y_0}{2y_0^2(1-y_0)^2}
(E_0-(1-y_0)K_0), \nonumber \\
E'(y_0) &=& -\frac{K_0-E_0}{2y_0}, \;
E''(y_0) = \frac{K_0-E_0}{4y_0^2}- \frac{E_0-(1-y_0)K_0}{4y_0^2(1-y_0)},
\end{eqnarray}
where $K_0 \equiv K(y_0), E_0 \equiv E(y_0)$. The leading term in the
expansion gives the momentum fraction $J_2/J$ in (\ref{jyj}) and
the subleading terms of order $1/\mt^2$ and 
order  $1/\mt^4$ respectively yield
\begin{eqnarray}
y_1 &=& -\frac{4y_0(1-y_0)K_0^2(K_0-E_0)(E_0-(1-y_0)K_0)}
{(E_0^2-(1-y_0)K_0^2)\pi^2},
\label{yon} \\
y_2 &=& \frac{(1-y_0)(K_0-E_0)(E_0-(1-y_0)K_0)}{E_0^2-(1-y_0)K_0^2}
\left( y_1^2 A + \frac{4y_1K_0^2}{\pi^2}\right)
\label{ytw}\end{eqnarray}
with
\begin{eqnarray}
A &=& \frac{2}{E_0-(1-y_0)K_0}\left( \frac{y_0K_0''}{2} + K_0'
\right) - \frac{y_0K_0(K_0''-E_0'')}{(K_0-E_0)(E_0-(1-y_0)K_0)}
\nonumber \\
&+& \frac{3}{4}\left\{ \frac{1}{(1-y_0)^2}\left(\frac{E_0}{K_0-E_0}
\right)^2 - \left(\frac{K_0}{E_0-(1-y_0)K_0}\right)^2 \right \}.
\end{eqnarray} 
To make the expansion of the eq. (\ref{tre}) multiplied by
$1/\mt^2$ we use (\ref{yon}), (\ref{ytw}) and (\ref{jyj})
to see that the $1/\mt^4$ term gives
\begin{equation}
\epsilon_2 = \frac{2K_0^3}{\pi^4}[(1-y_0)K_0 - (2-y_0)E_0].
\label{etw}\end{equation}
In ref. \cite{SS} the two transcendental equations (\ref{tre}) and 
(\ref{trt}) for the circular string were rewritten in 
Lagrange-inversion form, from which the same solution as (\ref{etw}) was
derived. Thus we have observed that the subleading energy correction
$E_2$ in (\ref{etk}) precisely agrees with the second-order energy 
correction $\epsilon_2\la^2/J^3$ in the classical circular string solution
with two $S^5$ spins $(J_1,J_2)$.

\section{The folded two-spin $(S,J)$ string solution}

Let us consider the general string configuration with two $AdS_5$ spins
$(S_1,S_2)$ and one $S^5$ spin $J$. From the classical string action
in $AdS_5 \times S^5$ the effective two-dimensional sigma model action
was constructed \cite{ST}. The relevant metric is expressed as
\begin{equation}
ds^2 = dY_i^*dY^i + d\alpha^2,
\end{equation}
where $\alpha$ is an angle in $S^5$ and the 3 complex coordinates
$Y_i \;(i=0,1,2)$ parametrize $AdS_5$ as
\begin{equation}
Y_0=\cosh\rho e^{it}, \; Y_1=\sinh \rho \cos\theta e^{i\phi_1}, \;
 Y_2=\sinh \rho \sin\theta e^{i\phi_2}
\end{equation}
and $Y^i=\eta^{ij}Y_j$ with $\eta^{ij} = \mathrm{diag}(-1,1,1)$ and
$Y_iY^i= -1$. The overall phase $y$ is introduced by $Y_i=e^{iy}V_i$
where $V_i$ are the coordinates of a non-compact version of $CP^2$.
By changing the coodinates as $u\equiv \alpha, v \equiv y- \alpha$ and
choosing the conformal gauge supplemented by the condition 
$u =\alpha = \mt \tau$ which fixes the residual conformal diffeomorphism
freedom, we have the following leading-order effective sigma model action
\begin{eqnarray}
S &=& J\int d\mathrm{t}\int_0^{2\pi}\frac{d\sigma}{2\pi} \left\{
-(B_\mathrm{t} + \pa_\mathrm{t}v) - \frac{1}{2}
D_{\sigma}^*V_i^* D_{\sigma}V^i \right \}, \nonumber \\
B_a &\equiv& iV_i^*\pa_aV^i\; (a=\mathrm{t}, \sigma), \hspace{1cm}
D_{\sigma}V_i =\pa_{\sigma}V_i - iB_{\sigma}V_i,
\label{sbv}\end{eqnarray}
where the time coordinate has been rescaled as $\tau = \mt \mathrm{t}$. 
The first term $B_\mathrm{t}$
plays the role of a Wess-Zumino term and here we leave the total 
derivative term $\pa_\mathrm{t}v$. The ``longitudinal" coordinate $v$ is
redundant but plays a role when it comes to the conformal gauge 
constraints. 

Now we focus on the two-spin SL(2) sector with one $AdS_5$ spin
$(S_1=S,S_2=0)$ and one $S^5$ spin $J$, whose relevant metric is
given by
\begin{equation}
ds^2 = -\cosh^2\rho dt^2 + d\rho^2 + \sinh^2\rho d\phi_1^2 + d\alpha^2.
\label{mgl}\end{equation}
We choose 
\begin{equation}
t = y + \eta, \hspace{1cm} \phi_1 = y - \eta, \hspace{1cm} \theta = 0
\label{tra}\end{equation}
to have $V_i = (\cosh\rho e^{i\eta}, \sinh\rho e^{-i\eta}, 0)$ and
\begin{equation}
ds^2 =  -(dy + B)^2 + d\alpha^2 + d\rho^2 + \sinh^22\rho d\eta^2,
\; B = \cosh 2\rho d\eta,
\label{dsy}\end{equation}
from which $y$ is regarded as a time coordinate of the target space. 
The leading-order 
effective sigma model action (\ref{sbv}) becomes
\begin{equation}
S = J\int d\mathrm{t} \int_0^{2\pi} \frac{d\sigma}{2\pi}\left \{ 
-\cosh 2\rho \pa_{\mathrm{t}}\eta - \frac{1}{2}[(\pa_{\sigma}\rho)^2 +
\sinh^2 2\rho(\pa_{\sigma}\eta)^2] \right \},
\end{equation}
where the total derivative term is here dropped. This sigma model action
was also produced from the SL(2) coherent state expectation value of
the Hamiltonian of the integrable $\mathrm{XXX}_{-1/2}$ spin chain
in the gauge theory side \cite{ST}. By rescaling the time coordinate
$\mathrm{t} \rightarrow \tilde{\la}\mathrm{t}$ and using 
the space coordinate $x = J\sigma/2\pi$, we have
\begin{equation}
S = \int d\mathrm{t} \int_0^{J} dx\left \{ -\cosh 2\rho 
\pa_{\mathrm{t}}\eta - \frac{\la}{8\pi^2}[(\pa_x\rho)^2 + 
\sinh^2 2\rho(\pa_x\eta)^2] \right \}.
\label{sco}\end{equation}
The equations of motion read
\begin{eqnarray} 
-2\sinh 2\rho \pa_{\mathrm{t}}\eta &+& \frac{\la}{4\pi^2}[\pa_x^2\rho
- 2\sinh 2\rho\cosh 2\rho(\pa_x\eta)^2 ] = 0,
\label{eqr} \\
-2\sinh 2\rho \pa_{\mathrm{t}}\rho &+& \frac{\la}{4\pi^2}\pa_x(
\sinh^2 2\rho\pa_x\eta) = 0
\label{eqe}\end{eqnarray}
with the boundary conditions 
\begin{equation}
\eta(x=J, \mathrm{t}) = \eta(x=0, \mathrm{t}),  \hspace{1cm} 
\rho(x=J, \mathrm{t}) = \rho(x=0, \mathrm{t}).
\label{boc}\end{equation}
We start to make an ansatz $\pa_x\eta = 0$ obeying (\ref{boc}).
The eq. (\ref{eqe}) yields $\pa_{\mathrm{t}}\rho = 0$. 
Then the derivative of (\ref{eqr}) with
respect to $\mathrm{t}$ gives $\pa_{\mathrm{t}}^2\eta = 0$, 
from which we have $\pa_{\mathrm{t}}\eta = \omega_{\eta}$. 
The eq. (\ref{eqr}) turns out to be
\begin{equation}
\pa_x^2\rho - \frac{8\pi^2\omega_{\eta}}{\la} \sinh 2\rho = 0,
\end{equation}
whose first integral is given by $(\pa_x\rho)^2/2 - 
(4\pi^2\omega_{\eta}/\la)\cosh 2\rho = c$. Thus we obtain
\begin{equation}
(\pa_x\rho)^2 = a - b\cosh 2\rho, \hspace{1cm} a = 2c, \; b = 
- \frac{8\pi^2\omega_{\eta}}{\la}.
\label{fir}\end{equation}
Here we consider the $\omega_{\eta}<0$ case which corresponds to
$b > 0$ as well as $c > 0$, that is $a > 0$. When $a > b$,
the folded string configuration is formed such that the maximal
value of $\rho$ is determined by $a - b \cosh 2\rho_0 = 0$.

Therefore the conserved momentum $P_{\eta}$ 
conjugate to $\eta$ is specified by
\begin{equation}
P_{\eta} = -\int_0^J dx \cosh 2\rho = -4\int_0^{\rho_0} d\rho
\frac{1 + 2\sinh^2\rho}{\sqrt{a - b\cosh 2\rho}},
\end{equation}
while the angular momentum associated with $u \equiv \alpha$ or the 
length of the spin chain is expressed as
\begin{equation}
J = \int_0^J dx = 4\int_0^{\rho_0}d\rho\frac{1}{\sqrt{a - b\cosh 2\rho}}.
\end{equation}
The integrations can be performed in 
terms of complete elliptic integrals as
\begin{eqnarray}
P_{\eta} &=& - \frac{4}{\sqrt{2b}}(2E(x)-K(x)) = -\frac{8}{\sqrt{2b}}
(E(x) - K(x)) - J, \nonumber \\
J &=& \frac{4}{\sqrt{2b}}K(x), \hspace{1cm} x = -\sinh^2\rho_0 =
- \frac{a - b}{2b}.
\label{pjx}\end{eqnarray}
Since $v$ is interpreted as a time coordinate and $i\pa_v = i\pa_y$,
the one-loop energy provided from (\ref{sco}) is expressed as 
\begin{equation}
E_v^{(1)} = \frac{\la}{8\pi^2}\int_0^J dx[(\pa_x\rho)^2 + 
\sinh^2 2\rho(\pa_x\eta)^2]
\label{evo}\end{equation}
and the energy in the coordinates of (\ref{dsy}) is given by
\begin{equation}
E_y = E_v = J + E_v^{(1)}.
\label{eyj}\end{equation}
When the solution is substituted into (\ref{evo}), we have
\begin{equation}
E_v^{(1)} = \frac{\la}{2\pi^2}\int_0^{\rho_0} d\rho
\frac{a - b\cosh 2\rho}{\sqrt{a - b\cosh 2\rho}} = \frac{\la}{8\pi^2}
(aJ + bP_{\eta}).
\label{ejp}\end{equation}
The conformal dimension of the gauge invariant 
operator is identified with the
energy in the global coordinates in $AdS_5$ whose relevant metric is here
described by (\ref{mgl}), so that the $AdS_5$ energy is given by
$E_t = i\pa_t$ and the $AdS_5$ spin by $S = -i\pa_{\phi_1}$.
From the coordinate transformation in (\ref{tra}) we find the relations
between the relevant conserved charges
\begin{eqnarray}
E_y &=& i\pa_y = i(\pa_t + \pa_{\phi_1}) = E_t - S,
\label{yts} \\
-P_{\eta} &=& i\pa_{\eta} =  i(\pa_t - \pa_{\phi_1}) = E_t + S.
\label{pst}\end{eqnarray}
Combining (\ref{yts}) with (\ref{eyj}), (\ref{ejp}) and (\ref{pjx}) 
we obtain
\begin{equation}
E_t = J + S + \frac{2\la}{\pi^2}\frac{K(x)}{J}[(1-x)K(x) - E(x)].
\label{ets}\end{equation}
The substitution of (\ref{ets}) and (\ref{pjx}) into (\ref{pst})
gives  to the leading order
\begin{equation}
1 + \frac{S}{J} = \frac{E(x)}{K(x)}.
\label{ekx}\end{equation}
The obtained expressions (\ref{ets}) and (\ref{ekx}) agree with
the results in ref. \cite{BFS} where 
the folded two-spin $(S,J)$  string solution
in $AdS_5 \times S^5$ is shown to be related by an
analytic continuation to the folded two-spin $(J_1,J_2)$ string
solution and the system of two transcendental
equations specifying this configuration is solved by the expansion
procedure.

\section{The circular three-spin $(S_1,S_2,J)$ string solution}

Let us turn to the three-spin $(S_1,S_2,J)$ string configuration in the
relevant $AdS_5 \times S^5$ metric
\begin{equation}
ds^2 = -\cosh^2\rho dt^2 + d\rho^2 + \sinh^2\rho(d\theta^2 + 
\cos^2\theta d\phi_1^2 + \sin^2\theta d\phi_2^2) + d\alpha^2.
\end{equation}
Setting 
\begin{equation}
t = y + \eta, \hspace{1cm} \phi_1 = y - \eta, 
\hspace{1cm} \phi_2 = y + \phi
\label{typ}\end{equation}
 we have
\begin{eqnarray}
V_i &=& (\cosh\rho e^{i\eta},\; \sinh\rho \cos\theta e^{-i\eta},
\; \sinh\rho \sin\theta e^{i\phi}), \nonumber \\
B_{\mathrm{t}} &=& (\cosh^2\rho + \sinh^2\rho \cos^2\theta)
\pa_{\mathrm{t}}\eta - \sinh^2\rho\sin^2\theta\pa_{\mathrm{t}}\phi.
\end{eqnarray}
Through the rescaling of the time coordinate $\mathrm{t} 
\rightarrow \tilde{\la}\mathrm{t}$ the effective sigma model 
action (\ref{sbv}) is expressed in terms of $\sigma$ here as
\begin{eqnarray}
S &=& -\frac{J}{2\pi}\int d\mathrm{t} \int_0^{2\pi}d\sigma \left\{ 
(\cosh^2\rho + \sinh^2\rho\cos^2\theta)\dot{\eta} - 
\sinh^2\rho\sin^2\theta\dot{\phi}
+ \dot{v} \right\} \nonumber \\
&-& \frac{\la}{4\pi J}\int d\mathrm{t}\int_0^{2\pi}d\sigma \{
\rho'^2 - (\cosh^2\rho - \sinh^2\rho\cos^2\theta)\eta'^2 
+ \sinh^2\rho \theta'^2 \nonumber \\  
&+& \sinh^2\rho\sin^2\theta\phi'^2 +
[(\cosh^2\rho + \sinh^2\rho\cos^2\theta)\eta' - 
\sinh^2\rho\sin^2\theta\phi']^2 \},
\label{sph}\end{eqnarray}
where the dot and prime denote 
derivatives with respect to $\mathrm{t}$ and
$\sigma$. This action is invariant under constant shifts in $\eta$ and
$\phi$ so that the corresponding conserved angular momenta are given by
\begin{eqnarray}
P_{\eta} &=& -\frac{J}{2\pi}\int_0^{2\pi}d\sigma (\cosh^2\rho +
\sinh^2\rho\cos^2\theta ), \label{pet}\\
P_{\phi} &=& \frac{J}{2\pi}\int_0^{2\pi}d\sigma \sinh^2\rho
\sin^2\theta.
\label{pph}\end{eqnarray}
The effective Hamiltonian is written by
\begin{eqnarray}
H &=& \frac{\la}{4\pi J}\int_0^{2\pi}d\sigma \{
\rho'^2 - (\cosh^2\rho - \sinh^2\rho\cos^2\theta)\eta'^2 
+ \sinh^2\rho\theta'^2 \nonumber \\
&+& \sinh^2\rho\sin^2\theta\phi'^2  
+ [(\cosh^2\rho + \sinh^2\rho\cos^2\theta)\eta' - 
\sinh^2\rho\sin^2\theta\phi']^2 \}.
\label{hth}\end{eqnarray}

From the action (\ref{sph}) the equations of motion for $\eta, \phi,
\rho, \theta$ are obtained respectively by
\begin{eqnarray}
&\pa_{\mathrm{t}}(\cosh^2\rho + 
\sinh^2\rho \cos^2\theta) - \frac{\la}{J^2}
\{ \pa_{\sigma}[(\cosh^2\rho - \sinh^2\rho\cos^2\theta)\eta']
 \nonumber \\ &- \pa_{\sigma}[(\cosh^2\rho + \sinh^2\rho\cos^2\theta)
((\cosh^2\rho + \sinh^2\rho\cos^2\theta)\eta' - 
\sinh^2\rho\sin^2\theta\phi')]\} = 0, 
\label{aeq} \\ 
&\pa_{\mathrm{t}}(\sinh^2\rho \sin^2\theta) + \frac{\la}{J^2}
\{ -\pa_{\sigma}(\sinh^2\rho\sin^2\theta) \nonumber \\
&+ \pa_{\sigma}[\sinh^2\rho\sin^2\theta((\cosh^2\rho + 
\sinh^2\rho\cos^2\theta)\eta' - \sinh^2\rho\sin^2\theta\phi')]\} = 0,
\label{beq} \\
&(1 + \cos^2\theta)\dot{\eta} - \sin^2\theta\dot{\phi} + \frac{\la}{2J^2}
\{-\frac{\rho''}{\sinh\rho\cosh\rho} - \sin^2\theta \eta'^2
+ \theta'^2 + \sin^2\theta\phi'^2 \nonumber \\
&+ 2[ (\cosh^2\rho + \sinh^2\rho\cos^2\theta)\eta'
 - \sinh^2\rho\sin^2\theta
\phi']((1 + \cos^2\theta)\eta' - \sin^2\theta\phi')\}=0, 
\label{ceq} \\
&\sinh^2\rho(\dot{\eta} + \dot{\phi}) + \frac{\la}{2J^2}
\{ \frac{\pa_{\sigma}(\sinh^2\rho\theta')}{\sin\theta\cos\theta} 
+ \sinh^2\rho\eta'^2 - \sinh^2\rho\phi'^2 \nonumber \\
&+ 2[ (\cosh^2\rho + \sinh^2\rho\cos^2\theta)\eta' - \sinh^2\rho
\sin^2\theta\phi']\sinh^2\rho(\eta' + \phi') \} = 0.
\label{deq}\end{eqnarray}
In order to solve the involved equations of motion we make a simple
ansatz; $\rho = \rho_0$ and $\theta = \theta_0$ with $\rho_0$ and
$\theta_0$ constant, and $\eta' = k_1, \phi' = k_2, v' = p$
where $k_1, k_2, p$ are integers. Substituting this ansatz into the
equations of motion we see that (\ref{aeq}) and (\ref{beq}) are
trivially satisfied and (\ref{deq}) yields
\begin{equation}
\dot{\eta} + \dot{\phi} =  \frac{\la}{2J^2}\{ k_2^2 -k_1^2 
- 2[k_1(\cosh^2\rho_0 + \sinh^2\rho_0\cos^2\theta_0) - 
k_2\sinh^2\rho_0\sin^2\theta_0](k_1 + k_2) \}.
\label{phk}\end{equation}

It is important to note that the relations between the conserved charges
for the three-spin configuration corresponding to (\ref{yts}) and 
(\ref{pst}) are given by
\begin{eqnarray}
E_y &=& i\pa_y = i(\pa_t + \pa_{\phi_1} + \pa_{\phi_2}) = 
E_t - S_1 - S_2, \label{yss}\\
-P_{\eta} &=& i\pa_{\eta} = i(\pa_t - \pa_{\phi_1}) = E_t + S_1,  
\label{pes}\\
-P_{\phi} &=& i\pa_{\phi} = i\pa_{\phi_2} = -S_2,
\end{eqnarray}
where the coordinate transformation in (\ref{typ}) has been taken into
account and $S_1$ and $S_2$ are the spins coming from the rotations
in the $\phi_1$ and  $\phi_2$ directions. The spin $S_2$ is 
so identified with $P_{\phi}$ in (\ref{pph}) that we have
\begin{equation}
S_2 = J\sinh^2\rho_0\sin^2\theta_0.
\label{srt}\end{equation}
The eq. (\ref{yss}) combines with $E_y = J + E_v^{(1)}$
corresponding to (\ref{eyj}), and yields
\begin{equation}
E_t = J + S_1 + S_2 + E_v^{(1)},
\label{jss}\end{equation}
where the one-loop energy correction $E_v^{(1)}$ is evaluated from the
effective Hamiltonian (\ref{hth}). Collecting eqs. (\ref{pet}), 
(\ref{pes}) and (\ref{jss}) we obtain to the leading order
\begin{equation}
2S_1 + S_2 = J\sinh^2\rho_0(1 + \cos^2\theta_0),
\end{equation}
which gives $S_1 = J\sinh^2\rho_0\cos^2\theta_0$, that is compared with
(\ref{srt}). Combining (\ref{phk}) with (\ref{ceq}) substituted by
the ansatz we derive 
\begin{eqnarray}
\dot{\phi} &=& -\frac{\la}{2J^2} \left\{ k_1^2 - k_2^2 
+ 2\left[k_1 + \frac{2k_1S_1}{J} + \frac{(k_1 - k_2)S_2}{J} 
\right]k_2 \right\}, \\
\dot{\eta} &=& -\frac{\la}{J^2}
\left[k_1 + \frac{2k_1S_1}{J} + \frac{(k_1 - k_2)S_2}{J} \right]k_1.
\end{eqnarray}
The one-loop energy correction is also evaluated as
\begin{equation}
E_v^{(1)} = \frac{\la}{2J^3}\{ -k_1^2J(J + S_2) + k_2^2JS_2
+ [k_1(J + 2S_1 + S_2) - k_2S_2 ]^2 \}.
\label{ekk}\end{equation} 
Since the momentum along the $\sigma$-direction should vanish, we find 
from the conformal gauge constraint that
\begin{equation}
P_{\sigma} = - \frac{\la}{2\pi} \int d\sigma [(\cosh^2\rho + 
\sinh^2\rho\cos^2\theta)\pa_{\sigma}\eta - \sinh^2\rho\sin^2\theta
\pa_{\sigma}\phi + \pa_{\sigma}v ] = 0,
\end{equation}
which yields 
\begin{equation}
2k_1S_1 + (k_1 - k_2)S_2 + (p + k_1)J = 0.
\label{lin}\end{equation}
If we identify $2k_1 = n_1,\; k_1 - k_2 = n_2, \; p + k_1 = m$
and take account of (\ref{lin}),
the clumsy expression (\ref{ekk}) turns out to be a symmetric one
\begin{equation}
E_v^{(1)} = \frac{\la}{2J^2}( m^2J + \sum_{a=1}^2 n_a^2S_a)
\end{equation}
with $mJ + \sum_{a=1}^2 n_aS_a = 0$. Thus we recover the result of 
ref. \cite{ART}, where the general circular constant-radii solution 
in $AdS_5 \times S^5$ is constructed by reducing the $AdS_5 \times S^5$
classical string theory to the Neumann-Rosochatius integrable system.

\section{Conclusion}

By constructing the explicit solution to the SU(2) spin chain sigma model
and reproducing precisely the leading and subleading energy corrections
of the circular string solution with two spins $(J_1,J_2)$ in the 
$AdS_5 \times S^5$ string theory, we have demonstrated the validity of the
one-loop \cite{MK} and two-loop \cite{KRT} expressions of the SU(2) spin
chain sigma model action.

For the SL(2) spin chain sigma model we have made an ansatz to solve its
equations of motion, and shown that this solution reproduces the
leading enegy correction of the folded string solution with two spins
$(S,J)$. From the general effective sigma model action we have recovered
the leading energy correction of 
the circular constant-radii string solution
with three spins $(S_1,S_2,J)$. Its involved equations of motion are
solved by a simple ansatz and the obtained energy expression in terms of
the three spins is out of order at the first glance.
However, by performing an appropriate redefinition of the relevant
parameters of the solution and taking account of the conformal gauge
constraint, we can extract a simple and symmetric energy expression.
Since the energy of the classical string solution in the global
coordinates in $AdS_5$ can be identified with the conformal dimemsion
of the composite operator in the gauge theory, we have observed that
there appears a subtlety in extracting the relevant energy from
the solution of the effective sigma model for both $(S,J)$ and
$(S_1,S_2,J)$ sectors, while  the $(J_1,J_2)$ sector does not include 
such subtlety. The subtlety is connected with the choice of the
time coordinate of the target space and resolved by making use of the
relations between the relevant conserved charges associated with
the coordinate transformation. Although there is such difference,
the folded $(S,J)$ sector exhibits some resemblance to the folded
$(J_1,J_2)$ sector as expected from the analytic continuation arguement.

Thus by making simple ans\"atze we have constructed several
particular classical solutions to the effective sigma models in order
to confirm the idea that there should exist the effective 
two-dimensional sigma model interpolating between the quantum planar
SYM theory and the classical string theory in $AdS_5 \times S^5$.
It would be interesting to search for the other type of solution
by making the different kind of ansatz for the $(S_1,S_2,J)$ sector.

While this paper being written, we learned of the work 
\cite{BCM} where the SL(2) spin chain sigma model was constructed
by using an approach different from \cite{HL} and making a
different coordinate transformation, and
the folded two-spin $(S,J)$ string solution was reproduced.

\end{document}